\RequirePackage[mathlines]{lineno}
\documentclass[amsmath,amssymb,twocolumn,prd,floatfix,showpacs, nofootinbib]{revtex4-1}
\usepackage{tabularx}  % tabular environment
\usepackage{bm, color} % math bold and /color
\usepackage{overpic,subfigure} % figures
\usepackage{multirow}
\usepackage{array}
\usepackage{dcolumn} % to deal with numbers with units and uncertainties
\usepackage[symbol]{footmisc}
\usepackage{epstopdf}
\usepackage{ulem,float}
\RequirePackage{xspace,cuted}

\newcommand{\gev}{\ensuremath{\mathrm{\,Ge\kern -0.1em V}}\xspace}
\newcommand{\mev}{\ensuremath{\mathrm{\,Me\kern -0.1em V}}\xspace}
\newcommand{\mevcc}{\ensuremath{{\mathrm{\,Me\kern -0.1em V\!/}c^2}}\xspace}

\def\chic#1{\ensuremath{\chi_{c#1}}\xspace} % dbm

\def\fz#1       {\ensuremath{f_0({#1})}\xspace}

\usepackage{lineno}
\usepackage{hyperref}
\hypersetup{colorlinks = true,
            linkcolor = black,
            urlcolor = blue,
            citecolor = blue}

\begin{document}

%\linenumbers

\title{\boldmath Observation of $\chi_{cJ}\to 3K_S^0K^\pm\pi^\mp$ }

\author{
M.~Ablikim$^{1}$, M.~N.~Achasov$^{4,c}$, P.~Adlarson$^{76}$, X.~C.~Ai$^{81}$, R.~Aliberti$^{35}$, A.~Amoroso$^{75A,75C}$, Q.~An$^{72,58,a}$, Y.~Bai$^{57}$, O.~Bakina$^{36}$, Y.~Ban$^{46,h}$
% ...
\\
\vspace{0.2cm}
(BESIII Collaboration)\\
\vspace{0.2cm} {\it
	$^{1}$ Institute of High Energy Physics, Beijing 100049, People's Republic of China\\
% ...
$^{4}$ Budker Institute of Nuclear Physics SB RAS (BINP), Novosibirsk 630090, Russia\\
% ...
$^{35}$ Johannes Gutenberg University of Mainz, Johann-Joachim-Becher-Weg 45, D-55099 Mainz, Germany\\
	$^{36}$ Joint Institute for Nuclear Research, 141980 Dubna, Moscow region, Russia\\
% ...
$^{46}$ Peking University, Beijing 100871, People's Republic of China\\
% ...
$^{57}$ Southeast University, Nanjing 211100, People's Republic of China\\
	$^{58}$ State Key Laboratory of Particle Detection and Electronics, Beijing 100049, Hefei 230026, People's Republic of China\\
% ...
$^{72}$ University of Science and Technology of China, Hefei 230026, People's Republic of China\\
% ...
$^{75}$ University of Turin and INFN, (A)University of Turin, I-10125, Turin, Italy; (B)University of Eastern Piedmont, I-15121, Alessandria, Italy; (C)INFN, I-10125, Turin, Italy\\
	$^{76}$ Uppsala University, Box 516, SE-75120 Uppsala, Sweden\\
% ...
$^{81}$ Zhengzhou University, Zhengzhou 450001, People's Republic of China\\
	\vspace{0.2cm}
	$^{a}$ Deceased\\
% ...
$^{c}$ Also at the Novosibirsk State University, Novosibirsk, 630090, Russia\\
% ...
$^{h}$ Also at State Key Laboratory of Nuclear Physics and Technology, Peking University, Beijing 100871, People's Republic of China\\
% ...
}
}

\begin{abstract}
  By analyzing $(2712.4\pm14.3)\times10^6$ $\psi(3686)$ events collected with the BESIII detector operating at the BEPCII collider, the decays $\chi_{c0,1,2} \to 3K_S^0K^\pm\pi^\mp$ are observed for the first time with statistical significances greater than $10\sigma$.
  The branching fractions of these decays are determined to be
  $\mathcal{B}(\chi_{c0}\to 3K_S^0K^\pm\pi^\mp )=(7.95\pm0.50\pm0.65)\times10^{-5},$
  $\mathcal{B}(\chi_{c1}\to 3K_S^0K^\pm\pi^\mp)=(2.62\pm0.08\pm0.19)\times10^{-4},$ 
  and
  $\mathcal{B}(\chi_{c2}\to 3K_S^0K^\pm\pi^\mp)=(1.72\pm0.07\pm0.15)\times10^{-4},$
  where the first uncertainties are statistical and the second systematic.
\end{abstract}

%\pacs{13.25.Gv, 14.40.Pq, 13.20.Gd}% PACS, the Physics and Astronomy Classification Scheme.
\maketitle

\section{Introduction}

The experimental studies of hadronic decays of the charmonium states are crucial for understanding Quantum Chromodynamics (QCD) models and QCD-based calculations~\cite{bes3-white-paper, qcd1, qcd2}, since charmonium physics lies in the transition region between pertubative and non-pertubative QCD. At present, both experimental and theoretical studies of hadronic decays of $\chi_{cJ}(J=0,1,2)$ states are 
not as extensive as for the vector charmonium states $J/\psi$ and $\psi(3686)$ (almost 30\% of the $\chi_{cJ}$ decays are not observed based on the total width), and more measurements are needed.
Since the $\chi_{cJ}$ mesons are produced copiously in the $\psi(3686)$ radiative decays, with branching fractions of about 9\%,
the world's largest $\psi(3686)$ data sample at BESIII~\cite{ref::psip-num-inc} offers good opportunities to study their different decays.
%Usually, the $\chi_{cJ}$ decays are studied via the $\psi(3686)\to \gamma\chi_{cJ}$ processes, which have large branching fraction of about 9\%~\cite{ref::pdg2024}. However, 
Currently, there have been theoretical predictions and experimental measurements of multi-body-meson decays of $\chi_{cJ}$~\cite{2004kpm,2010cs,ref::pdg2024}; 
however, the information of five-body hadronic $\chi_{cJ}$ decays is still very limited, 
and additional experimental measurements of these decays are needed.
% to deeply understand the decay dynamics of charmonium states.

In 2015, BESIII reported the measurement of the branching fractions of $\chi_{cJ} \to \bar{K}^0 K^\pm \pi^\mp\phi$ and $\chi_{cJ} \to K^+ K^- \pi^0\phi$ decays in the final state $\chi_{cJ} \to KKKK\pi$~\cite{ref::pdg2024,ref::kkpiphi}, by using $106 \times 10^6$ $\psi(3686)$ events.
The obtained results are listed in Table~\ref{tab::reledecay}. 
In the limit of isospin symmetry, the branching fractions of the $\chi_{cJ}\to 3K_S^0K^\pm\pi^\mp$ decays are expected to 
be at $10^{-4}$ level, which are accessible at BESIII but have not been studied to date.
In this paper, we determine the branching fractions of $\chi_{cJ}\to 3K_S^0K^\pm\pi^\mp$ for the first time by analyzing $(2712.4\pm14.3)\times10^6$ $\psi(3686)$ events collected with the BESIII detector.

\begin{table}[H]
	\centering
	\caption{The world average values $(\times 10^{-4})$ of the branching fractions of the decays contributing to $\chi_{cJ} \to KKKK\pi$.}
	\begin{tabular}{c|c|c|c}
		\hline
		\hline
		&$\chi_{c0}$&$\chi_{c1}$&$\chi_{c2}$ \\
		\hline
		$\bar{K}^0 K^\pm \pi^\mp\phi$&$37\pm6$&$33\pm5$ &$48\pm7$ \\
		\hline
		$K^+ K^- \pi^0\phi$&$19.0\pm3.5$&$16.2\pm 3.0$ &$27\pm5$ \\ \hline
		\hline
	\end{tabular}
	\label{tab::reledecay}
\end{table}

%\setlength{\parskip}{1ex}
%In this paper, by analyzing $(27.12\pm0.14)\times10^8$ $\psi(3686)$ events~\cite{ref::psip-num-inc} collected with the BESIII detector~\cite{ref::BesIII}, we present the first observation and branching fraction measurements of $\chi_{cJ} \to 3(K^+K^-)$.

\section{BESIII DETECTOR AND MONTE CARLO SIMULATION}
\label{sec:BES}
The BESIII detector~\cite{ref::detector} records symmetric $e^+e^-$ collisions provided by the BEPCII storage ring~\cite{ref::collider}
in the center-of-mass energy range from 1.84 to 4.95~GeV,
with a peak luminosity of $1.1 \times 10^{33}\;\text{cm}^{-2}\text{s}^{-1}$
achieved at $\sqrt{s} = 3.773\;\text{GeV}$.

The cylindrical core of the BESIII detector covers 93\% of the full solid angle and consists of a helium-based
 multilayer drift chamber~(MDC), a plastic scintillator time-of-flight
system~(TOF), and a CsI(Tl) electromagnetic calorimeter~(EMC),
which are all enclosed in a superconducting solenoidal magnet
providing a 1.0~T magnetic field. The magnetic field was 0.9 T in 2012, which affects 12.7\% of the total $\psi(3686)$ data.
The solenoid is supported by an
octagonal flux-return yoke with resistive plate counter muon
identification modules interleaved with steel.
The charged-particle momentum resolution at $1~{\rm GeV}/c$ is
$0.5\%$, and the
${\rm d}E/{\rm d}x$
resolution is $6\%$ for electrons
from Bhabha scattering. The EMC measures photon energies with a
resolution of $2.5\%$ ($5\%$) at $1$~GeV in the barrel (end-cap)
region. The time resolution in the TOF barrel region is 68~ps, while
that in the end cap region was 110~ps.
The end-cap TOF system was upgraded in 2015 using multi-gap resistive plate chamber technology, providing a time resolution of 60~ps~\cite{Tof1,Tof2,Tof3}, which benefits 83.3\% of the data used in this analysis.

Simulated data samples produced with a {\sc
geant4}-based~\cite{Geant4} Monte Carlo (MC) package, which
includes the geometric description of the BESIII detector and the
detector response, are used to determine detection efficiencies
and to estimate backgrounds. The simulation includes the beam
energy spread and initial state radiation (ISR) in the $e^+e^-$
annihilations with the generator {\sc
kkmc}~\cite{Jadach01}.

The inclusive MC sample includes the production of the
$\psi(3686)$ resonance, the ISR production of the $J/\psi$, and
the continuum processes incorporated in {\sc
kkmc}~\cite{Jadach01}.
All particle decays are modelled with {\sc
evtgen}~\cite{Lange01,Lange02} using branching fractions
either taken from the
Particle Data Group (PDG)~\cite{ref::pdg2024}, when available,
or otherwise estimated with {\sc lundcharm}~\cite{Lundcharm00}.
Final state radiation~(FSR)
from charged final state particles is incorporated using the {\sc
photos} package~\cite{PHOTOS}.
An inclusive MC sample containing $2.747\times10^{9}$ generic $\psi(3686)$ events is used to study background.
To account for effects of intermediate
resonance structures on the efficiency, each of the $\chi_{c0,1,2}$ is modeled using 200,000 mixed MC events, in which the dominant decay modes containing resonances are included. Particle decays are generated by {\sc evtgen}~\cite{Lange01,Lange02} for the known decay modes with branching fractions taken from the PDG~\cite{ref::pdg2024}.  The other decays are generated by the phase-space (PHSP) model.
%The dominant decay modes containing the resonances $f(1500)$ and $K^*(892)$ are mixed with the phase-space (PHSP) in our signal MC samples.
The mixing ratios of resonances are determined by examining the corresponding invariant mass distributions as discussed in Section IV, which are shown in Table~\ref{tab:weight}.

%\begin{table}[H]
%	\centering
%	\caption{generators used in signal MC sample.}
%	\label{tab::generator}
%	\begin{tabular}{c|c}
%		\hline
%		\hline
%		Decay modes & generator \\ \hline
%		$\psi(3686)\to \gamma\chi_{cJ}$ & P2GCJ \\ \hline
%		\begin{tabular}{c}
%			$\chi_{cJ}\to3K_S^0K^\pm\pi^\mp$\\
%			$\chi_{cJ}\to2K_S^0K^*(892)^\pm K^\mp$ \\
%			$\chi_{cJ}\to3K_S^0K^*(892)^0+c.c.$ \\
%			$\chi_{cJ}\to f_0(1500)K_S^0K^\pm\pi^\mp$ \\
%			$f_0(1500)\to K_S^0K_S^0$ \\
%			$K_S^0\to \pi^+\pi^-$
%		\end{tabular}
%		 & PHSP\\ \hline
%		\begin{tabular}{c}
%		 $K^*(892)^\pm \to K^0\pi^\pm$ \\
%		 $K^*(892)^0 \to K^\pm\pi^\mp$
%		\end{tabular} & VSS\\
%		\hline
%		\hline
%	\end{tabular}
%\end{table}

\section{EVENT SELECTION}
\label{sec:selection}

Candidate events are reconstructed via the charmonium
transitions $\psi(3686)\to\gamma\chic{J}$ followed by the hadronic
decays $\chic{J}\to 3K_S^0K^\pm\pi^\mp$. They are required to contain at least eight charged tracks and at least one photon candidate.

 Charged tracks (not originating from  $K_S^0$ decays) are required to be within a polar angle ($\theta$) range of $|\rm{cos\theta}|<0.93$, where $\theta$ is defined with respect to the $z$ axis, which is the symmetry axis of the MDC. The distance of closest approach to the interaction point (IP)
must be less than 10\,cm along the $z$ axis, $|V_{z}|$, and less than 1\,cm in the transverse plane, $|V_{xy}|$.

Photon candidates are identified using showers in the EMC. The deposited energy of each shower must be more than 25~MeV in the barrel region ($|\cos \theta|< 0.80$) and more than 50~MeV in the end-cap region ($0.86 <|\cos \theta|< 0.92$). To exclude showers originating from charged tracks, the angle subtended by the EMC shower and the position of the closest charged track at the EMC must be greater than 10$^\circ$ as measured from the IP. To suppress electronic noise and showers unrelated to the event, the difference between the EMC time and the event start time is required to be within [0, 700]\,ns.

Particle identification~(PID) for charged tracks combines measurements of the specific energy loss (d$E$/d$x$) in the MDC and the flight time in the TOF to form likelihoods $\mathcal{L}(h)~(h=K,\pi)$ for each hadron hypothesis $h$.
Charged kaons and pions are identified by comparing the likelihoods for the kaon and pion hypotheses and requiring  $\mathcal{L}(K)>\mathcal{L}(\pi)$ and $\mathcal{L}(\pi)>\mathcal{L}(K)$, respectively.

\begin{figure}[H]
	\centering
	\includegraphics[width=8.5cm]{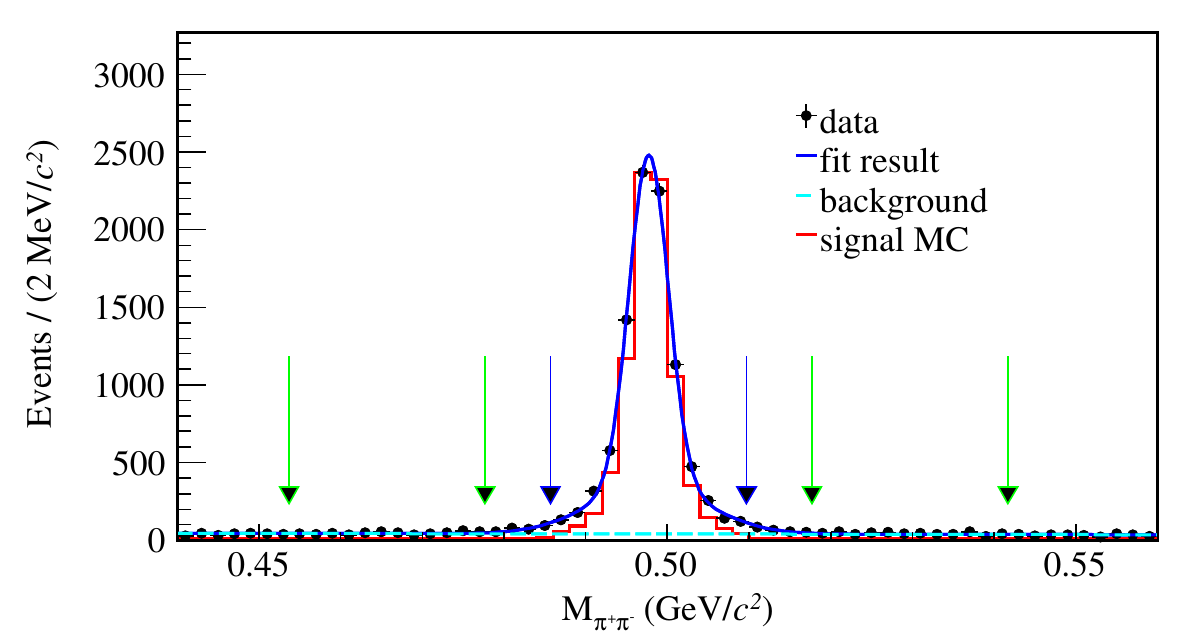}
	\caption{
		The $M_{\pi^+\pi^-}$ distribution of the accepted candidates. Dots with error bars represent the data, the blue line is the fit result, the red line is the signal MC sample and the green line the background. The signal MC sample is normalized to have the maximum height of the data. The pair of blue arrows shows the $K_S^0$ signal region, and the pairs of green arrows show the $K_S^0$ sideband regions.
	}
	\label{tab::ks1d}
\end{figure}

Each $K_{S}^0$ candidate is reconstructed from two oppositely charged tracks satisfying $|V_{z}|<$ 20~cm.
The two charged tracks are assigned as $\pi^+\pi^-$ without imposing further PID criteria. They are constrained to
originate from a common vertex. The
decay length of the $K^0_S$ candidate is required to be greater than
twice the vertex resolution away from the IP.	
The quality of the vertex fits is required to be  $\chi^2<200$.
The $K^{0}_{S}$ signal region is set to be within $3\sigma$, $|M_{\pi^+ \pi^-}-0.498|<0.012~\mathrm{GeV/}$${c}^2$, and the sideband regions are set to be within $5\sigma$ and $11\sigma$, $0.020<|M_{\pi^+ \pi^-}-0.498|<0.044~\mathrm{GeV/}$$c^2$. Figure ~\ref{tab::ks1d} shows the $M_{\pi^+\pi^-}$ distribution of the accepted candidates in data.

A four-constraint (4C) kinematic fit is performed under the hypothesis $\psi(3686)\to\gamma 3K_S^0K^\pm\pi^\mp$, by requiring the total reconstructed four-momentum to be conserved. In each event, if more than one combination survives, the one with the smallest $\chi_{\rm 4C}^{2}$ from the 4C kinematic fit is retained. The requirement on $\chi^2_{\rm 4C}$ is optimized with the Figure-of-Merit (FOM) defined as
\begin{equation}
\mathrm{FOM} = \frac{\mathit{S}}{\sqrt{\mathit{S}+\mathit{B}}},
\end{equation}
where $S$ is the number of events from the signal MC sample, normalized according to the pre-measured branching fractions without the FOM optimization,
and $B$ is the number of background events from the inclusive MC sample, normalized to the data size. From the optimization, we choose $\chi^2_{\rm 4C}<50$ as the nominal requirement.

%\section{Background Analysis}
%\label{sec:background}

The potential background components from $\psi(3686)$ decays are studied by analyzing the inclusive MC sample with a generic
event type analysis tool, TopoAna~\cite{topoana}. The peaking background in inclusive MC can be described by the $K_S^0$ sideband events in data, while the remaining background in inclusive MC is flat.
Furthermore, we examine the data sample taken at $\sqrt s =$ 3.650 GeV, corresponding to an integrated luminosity of 0.4~fb$^{-1}$~\cite{lum}. The background at this energy point is found to be negligible.

%\begin{itemize}
%	\item  Type a: events without $ \chi_{cJ}$
%	\item  Type b: events with $ \chi_{cJ}$, but without $ K^0_S$,which are possible come from $ \chi_{cJ} \to p K \Lambda$ process
%	\item  Type c: events with $ \chi_{cJ}$ and $ K^0_S$
%\end{itemize}

\section{Branching fraction}

\begin{figure}[H]
	\centering
	\includegraphics[width=8cm]{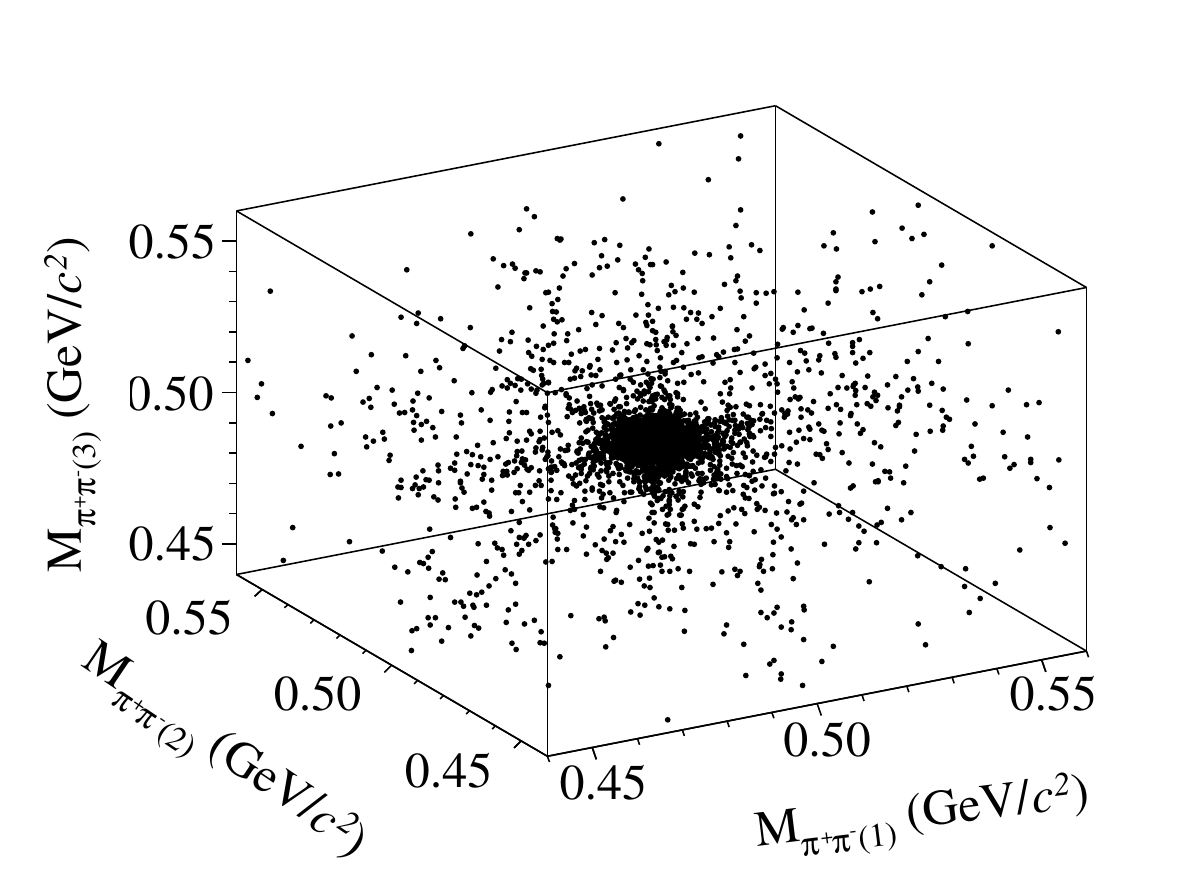}
	\caption{Distribution of $M_{\pi^+\pi^-(1)}:M_{\pi^+\pi^-(2)}:M_{\pi^+\pi^-(3)}$ for the candidate events in data.
		%The red, blue, yellow and green cubes denote the 3$K_S^0$ signal, SB1, SB2 and SB3 regions.
	}
	\label{fig::3dsideband}
\end{figure}

To determine the signal yields, a simultaneous unbinned maximum likelihood fit is performed on the $ M_{3K_S^0K^\pm\pi^\mp}$ distributions of the accepted candidate events in the $K^0_S$ signal and sideband regions. In this analysis, the combinatorial background in the $M_{\pi^+\pi^-}$ distribution is assumed to be flat. Thus, the net numbers of the $\chi_{cJ} \to 3K_S^0K^\pm\pi^\mp$ decays can be calculated by
\begin{equation}
	N_{\rm net}=N_{\rm sig}-\frac{1}{2}N_{\rm SB1}+\frac{1}{4}N_{\rm SB2}-\frac{1}{8}N_{\rm SB3},
\end{equation}
where $N_{\rm sig}$ is the number of events with three $\pi^+\pi^-$ pairs lying in the $K_S^0$ signal region, called the Signal region; $N_{\rm SB1}$ is the number of events with two $\pi^+\pi^-$ pairs lying in the $K_S^0$ signal region and the other one lying in the $K_S^0$ sideband region, called the SB1 region; $N_{\rm SB2}$ is the number of events with one $\pi^+\pi^-$ pair lying in the $K_S^0$ signal region and the other two lying in the $K_S^0$ sideband region, called the SB2 region; $N_{\rm SB3}$ is the number of events with three $\pi^+\pi^-$ pairs lying in the $K_S^0$ sideband region, called the SB3 region. Figure~\ref{fig::3dsideband} shows the three-dimensional distribution of the invariant mass of the three  $\pi^+\pi^-$ pairs in data. %$M_{\pi^+\pi^-(1)}$ versus $M_{\pi^+\pi^-(2)}$ versus $M_{\pi^+\pi^-(3)}$ of the candidate events in data.

\begin{figure}[H]
	\centering
	\includegraphics[width=4.27cm]{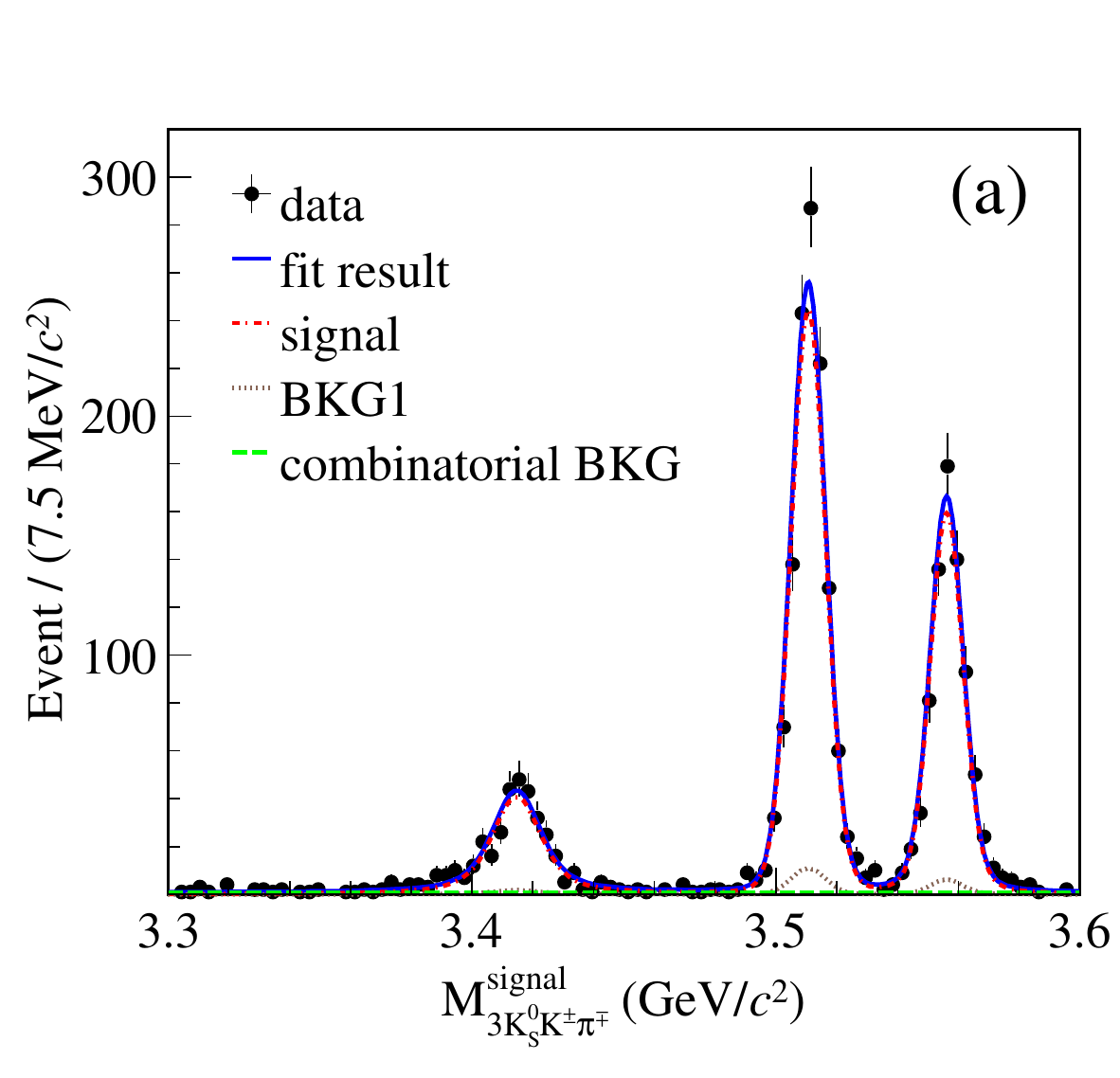}
	\includegraphics[width=4.27cm]{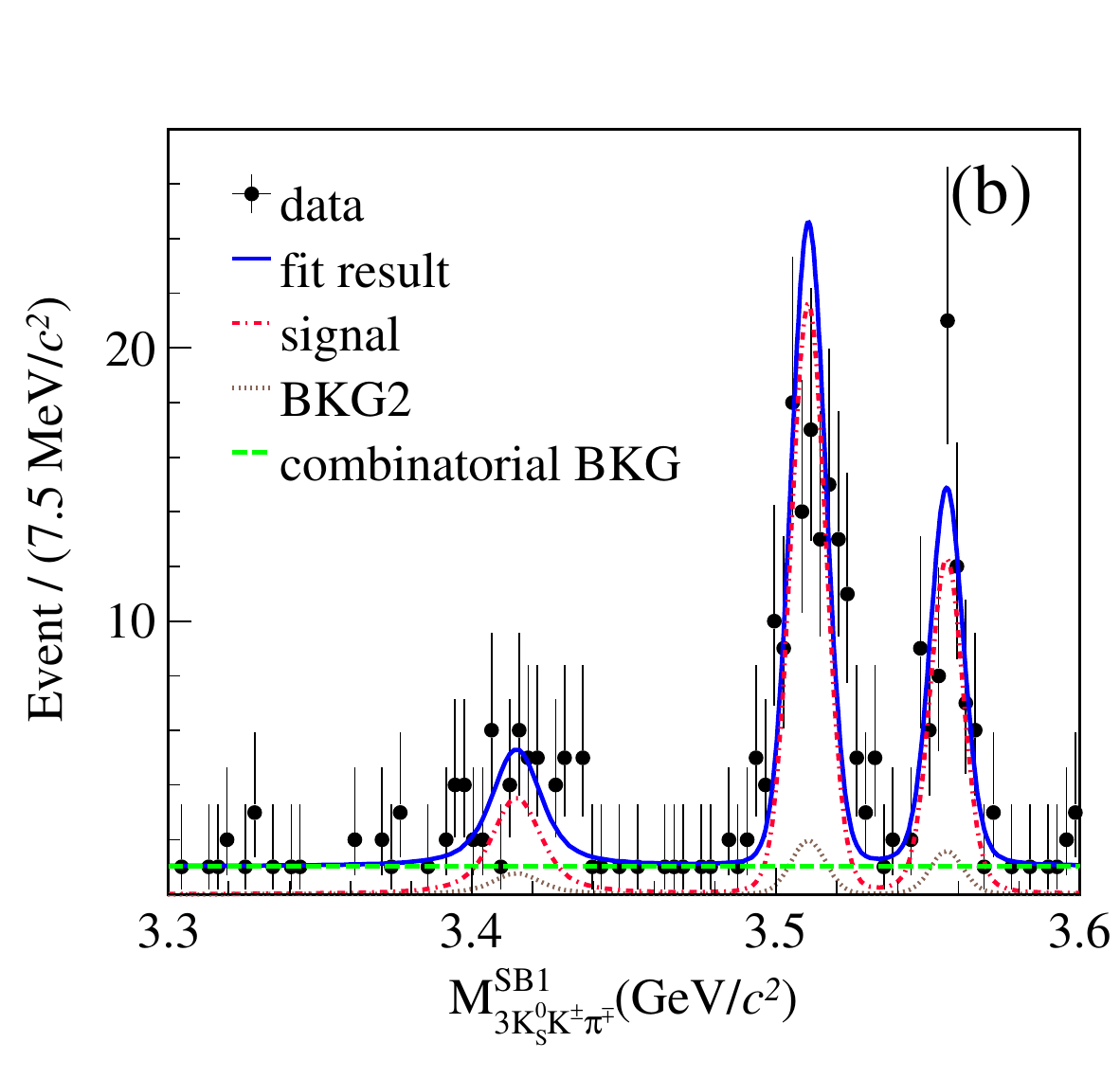}
	\includegraphics[width=4.27cm]{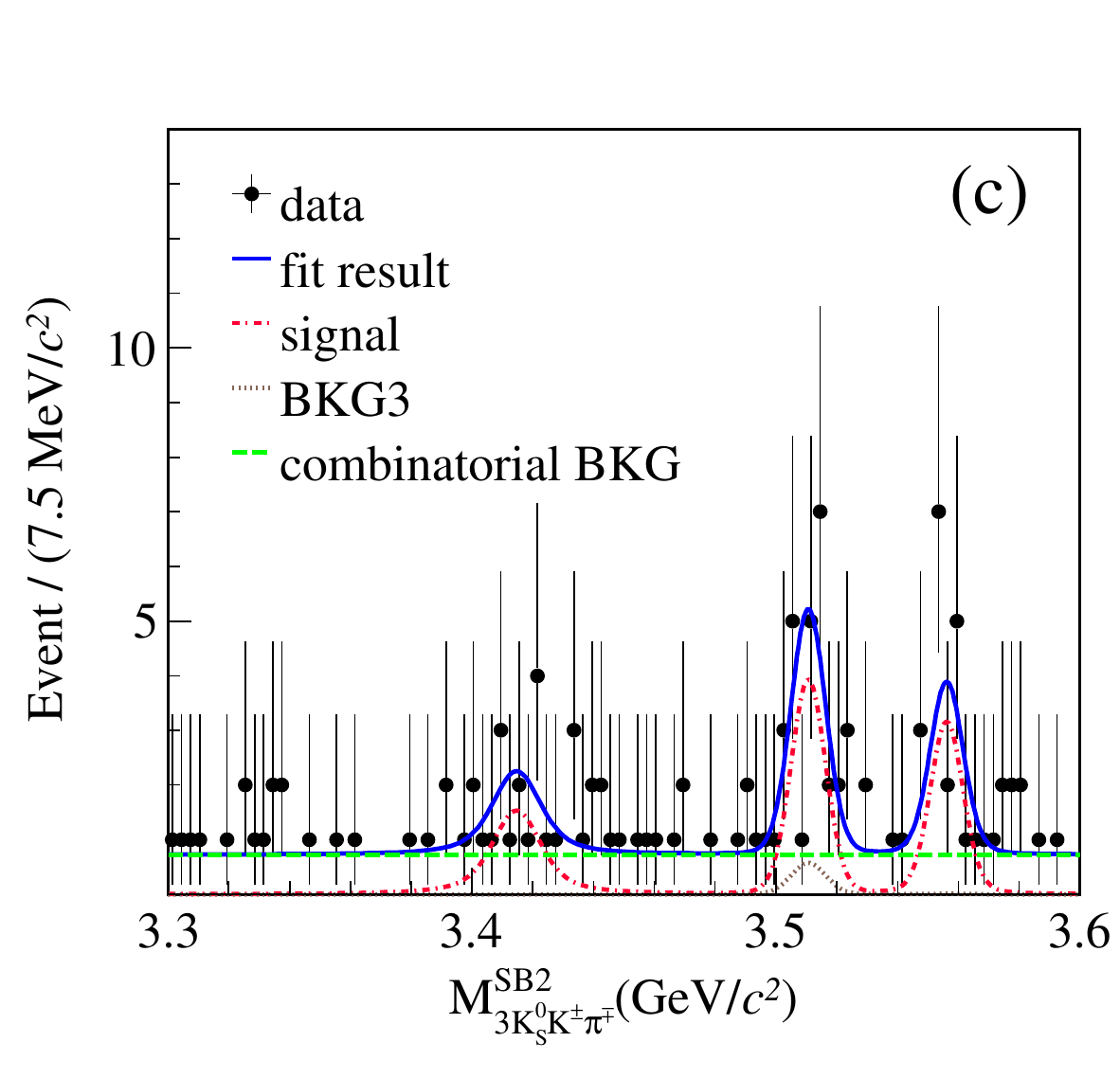}
	\includegraphics[width=4.27cm]{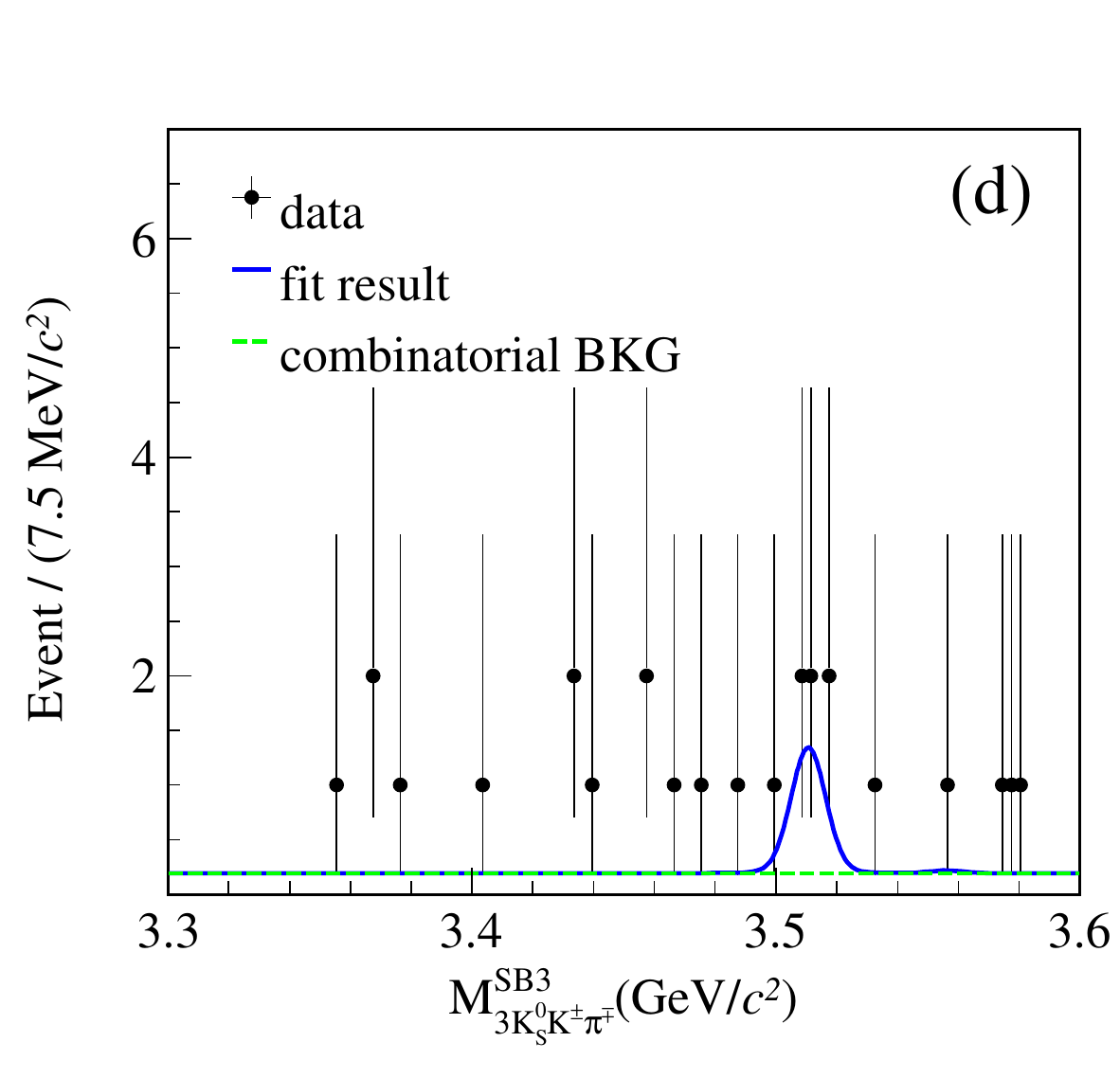}
	\caption{Simultaneous fit to the $M_{3K_S^0K^\pm\pi^\mp}$ distributions of the accepted candidate events for $\chi_{cJ}\to 3K_S^0K^\pm\pi^\mp$
		in data (dots with error bars) in the (a) Signal, (b) SB1, (c) SB2, and (d) SB3 regions, where BKG1, BKG2 and BKG3 stay for the fixed peaking background events from the fitted signal event candidates of sideband regions (b)(c) and (d). The blue solid curves are the fit results, the red curves are the signal shapes, the brown curves are the backgrounds constrained by the $K_S^0$ sidebands, and the green curves are the fitted backgrounds.}
	\label{tab:k-pi+fit_sig}
\end{figure}

 In the fit, the signal shape of the $\chi_{cJ}$ states is modeled by using a Breit-Wigner function convolved with a Gaussian function with free parameters, to take into account the resolution. The mass and width of each Breit-Wigner function are fixed to the respective values from the PDG. The backgrounds are described by a first-order Chebychev polynomial function for the events from both the $K_S^0$ signal and sideband regions. Figure \ref{tab:k-pi+fit_sig} shows the fit results. Significant $\chi_{cJ}$ signals are seen for the events in the signal region, while there are also some peaking background events from the sideband regions. The number of peaking background events in the signal region is estimated to be $\frac{1}{2}N_{\rm SB1}-\frac{1}{4}N_{\rm SB2}+\frac{1}{8}N_{\rm SB3}$, as defined before.
 %The backgrounds of SB1, SB2 and SB3 in (a), (b), (c) are fixed to the number of the fit results in (b), (c), (d).
 From this fit, the signal yields of $\chi_{c0}$, $\chi_{c1}$ and $\chi_{c2}$ decays, $N_{\chi_{cJ}}^{\rm obs}$, are obtained to be $343\pm22$, $1190\pm36$ and $767\pm29$, respectively. The statistical significances are larger than $10\sigma$ for all $\chi_{cJ}$ decays, which are estimated by comparing the likelihood values of the fit with and without each signal component and taking into account the number of degrees of freedom.

\begin{figure*}[htbp]
	\centering
	\includegraphics[width=0.98\linewidth]{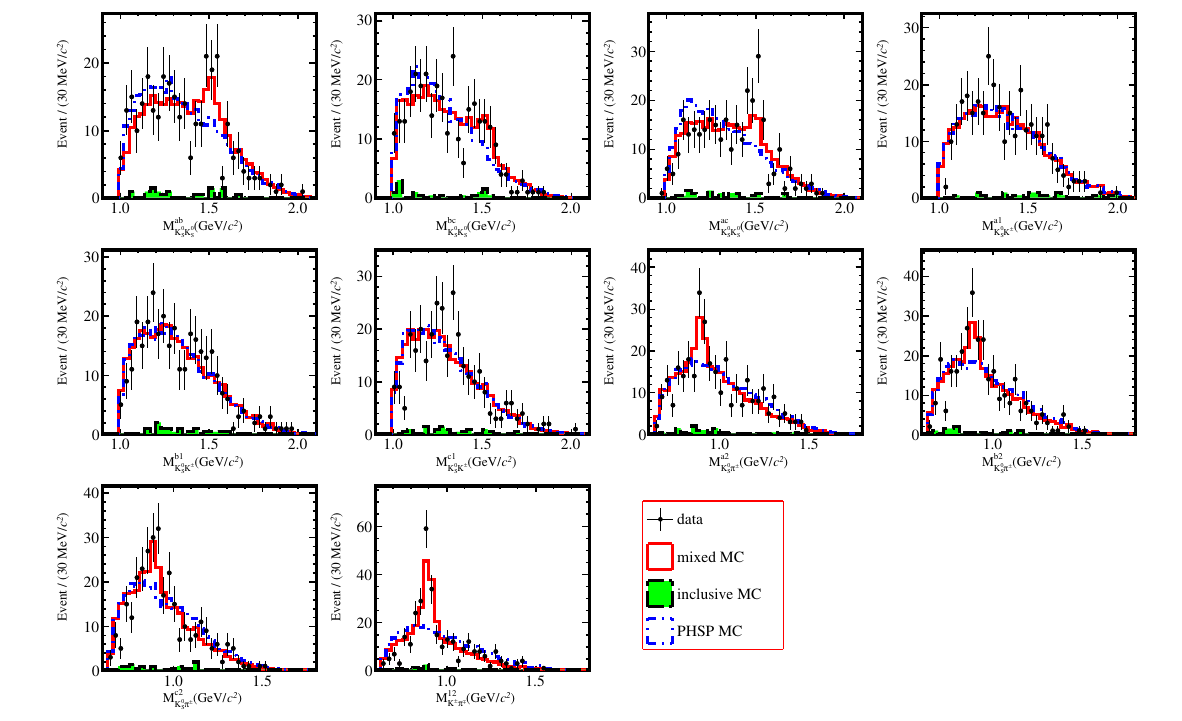}	
	\caption{Distributions of the two-body invariant masses of the accepted candidate events for $\chi_{c0}\to 3K_S^0K^\pm\pi^\mp$. The black dots with error bars are data, the red histograms are the mixed MC sample, the cyan histograms are the PHSP signal MC sample, and the green histograms are the simulated background from the inclusive MC sample. Events have been required to be within $M_{3K_S^0K^\pm\pi^\mp} \in[3.394, 3.434]$ GeV/$c^2$.}
	\label{tab:con0}
\end{figure*}

\begin{figure*}[htbp]
	\centering
	\includegraphics[width=0.98\linewidth]{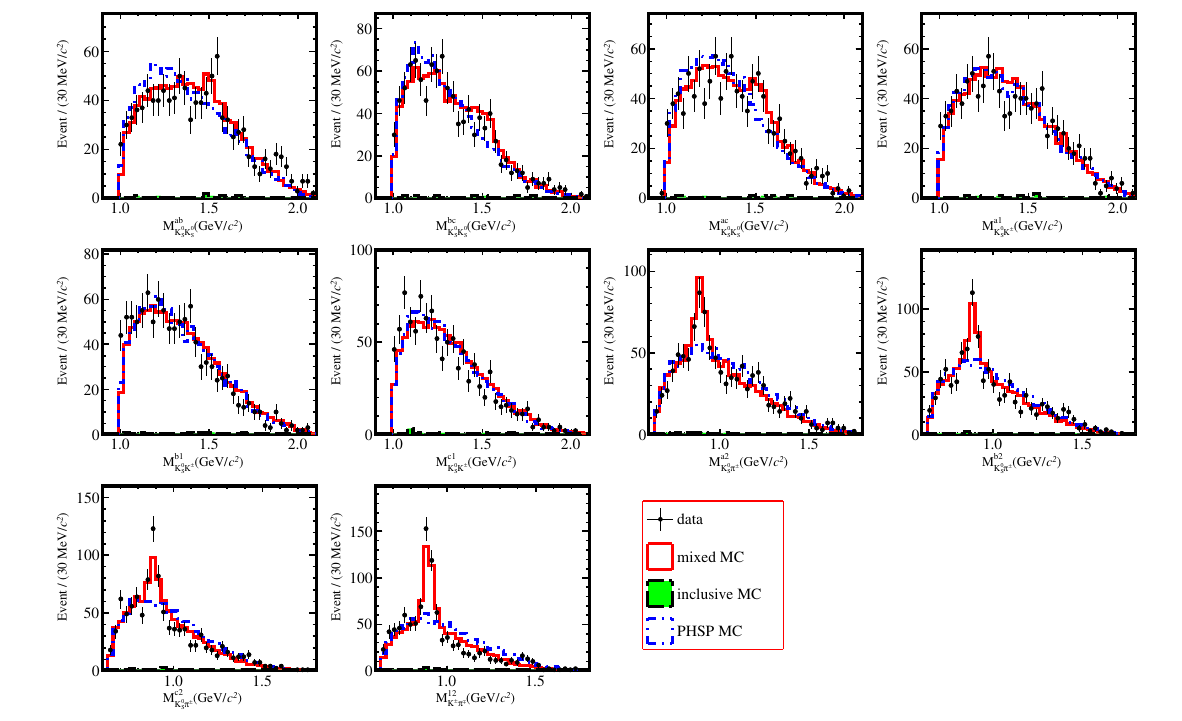}
	\caption{Distributions of the two-body invariant masses of the accepted candidate events for $\chi_{c1}\to 3K_S^0K^\pm\pi^\mp$. The black dots with error bars are data, the red histograms are the mixed MC sample, the cyan histograms are the PHSP signal MC sample, and the green histograms are the simulated background from the inclusive MC sample. Events have been required to be within $M_{3K_S^0K^\pm\pi^\mp} \in[3.496, 3.526]$ GeV/$c^2$.}
	\label{tab:con1}
\end{figure*}

\begin{figure*}[htbp]
	\centering
	\includegraphics[width=0.98\linewidth]{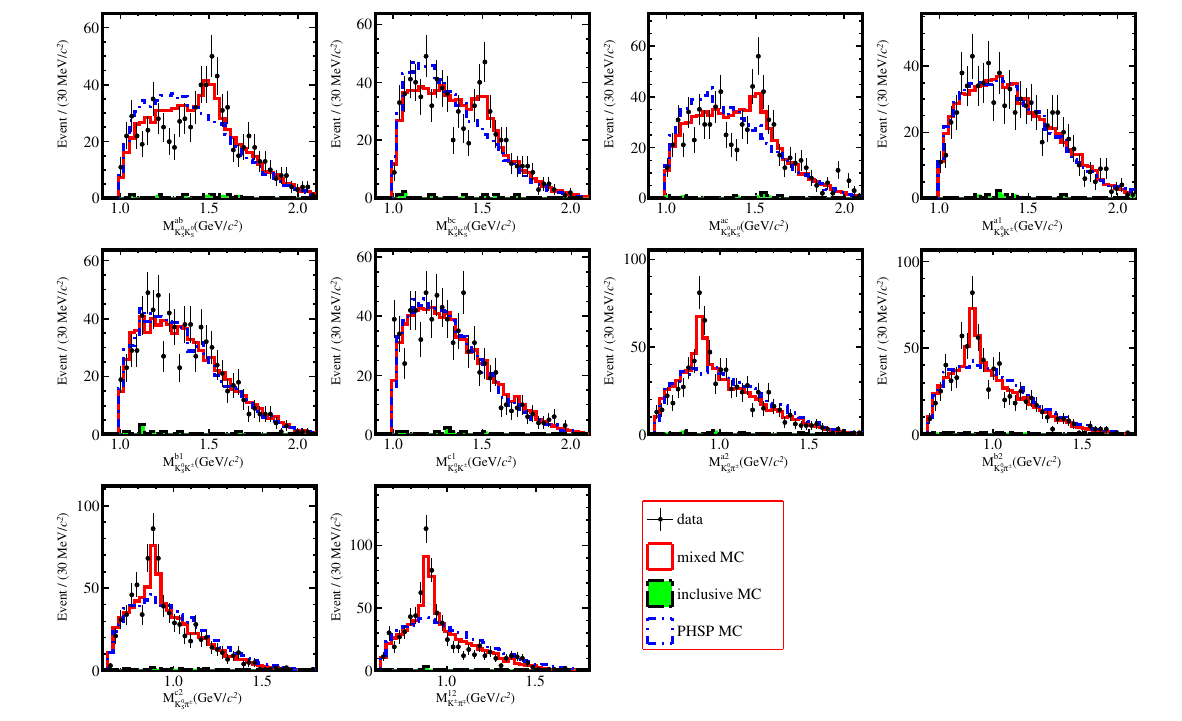}
	\caption{Distributions of the two-body invariant masses of the accepted candidate events for $\chi_{c2}\to 3K_S^0K^\pm\pi^\mp$. The black dots with error bars are data, the red histograms are the mixed MC sample, the cyan histograms are the PHSP signal MC sample, and the green histograms are the simulated background from the inclusive MC sample. Events have been required to be within $M_{3K_S^0K^\pm\pi^\mp} \in[3.541, 3.571]$ GeV/$c^2$.}
	\label{tab:con2}
\end{figure*}

The detection efficiencies of $\psi(3686)\to\gamma\chi_{cJ}$ with $\chi_{cJ}\to 3K_S^0K^\pm\pi^\mp$
are determined with the mixed MC samples with a certain fraction of
$\chi_{cJ}\to  3K_S^0K^\pm\pi^\mp$,
$\chi_{cJ}\to 2K_S^0K^*(892)^{\pm}K^\mp$, $\chi_{cJ}\to 3K_S^0\bar{K^*}(892)^0+c.c.$ and
$\chi_{cJ}\to f_X(1500)K_S^0K^\pm\pi^\mp$. The fractions are determined from fits to the invariant mass distributions of the two-body combinations, where the signal is modeled by a signal MC shape convolved with a Gaussian function and the background by a third-order Chebyshev polynomial, and are summarized in Table~\ref{tab:weight}. Figures \ref{tab:con0},~\ref{tab:con1} and \ref{tab:con2} show the comparisons of the invariant masses of the two-body combinations of the accepted candidate events for $\chi_{c0,1,2}\to 3K_S^0K^\pm\pi^\mp$ in data and in mixed MC samples. Here, we mark the three different $K_S^0$ candidates as a, b and c, and $K^\pm$ and $\pi^\pm$ as  1 and 2, respectively, to identify the different combinations.
The variations of these fractions will be considered as a source of systematic uncertainty.
\begin{table}[htbp]
	\centering
	
    \caption{The fractions (in \%) of different subprocesses in $\chi_{cJ}\to 3K_S^0K^\pm\pi^\mp$.} %, where the uncertainties are statistical only.}
	\begin{tabular}{l|ccc}
		
		\hline
		%		& \multicolumn{3}{c|}{$ p \bar p K^{0}_{S} K^{\mp} \pi^{\pm} $} \\
		\hline
		& \phantom{1}$\chi_{c0}$\phantom{1} & \phantom{1}$ \chi_{c1}$\phantom{1} & \phantom{1}$\chi_{c2}$\phantom{1} \\
		\hline
		
		$ 3K_S^0K^\pm\pi^\mp$(PHSP) & $3.1$ & $10.0$ & $4.7$ \\
		\hline
		
		$2K_S^0K^*(892)^{\pm}K^\mp$ & $33.0$ & $42.1$& $40.1$ \\
		\hline
		
		$ 3K_S^0\bar K^*(892)^{0}+c.c.$ & $31.3$ & $27.1$ &$22.1$\\
		\hline
		$ f_X(1500)K_S^0K^\pm\pi^\mp$ & $32.6$ & $20.8$ &$33.0 $\\
		\hline
		\hline
		
	\end{tabular}
	
	\label{tab:weight}
\end{table}

\label{sec:mc}

%For each decay of
%$\psi(3686)\to\gamma\chic{J}$, $\chic{J}\to 3(K^+K^-)$, about $10.8\times10^5$ signal MC events are generated
%using a $1+\lambda\cos^2\theta$ distribution, where $\theta$ is the angle between the radiative photon and beam directions, and $\lambda=1,-1/3,1/13$ for $J=0,1,2$ in accordance with
%the expectations for electric dipole transitions~\cite{ref::generate}.
%Intrinsic width and mass values in PDG~\cite{ref::pdg2024} are used to simulate the \chic{J} states.

The products of branching fractions of $\chi_{cJ}\to 3K_S^0K^\pm\pi^\mp$ and $\psi(3686)\to \gamma\chi_{cJ}$ are calculated as
%\begin{equation}
%	\centering
%	\begin{split}
%		{\mathcal B(\psi(3686)\to \gamma\chi_{cJ})}= \phantom{11111111111111111111111111}\\
%		\frac{N^{\rm net}_{\chi_{cJ}}}{N_{\psi(3686)}\cdot\varepsilon \cdot \mathcal{B}(\chi_{cJ}\to 3K_S^0K^\pm\pi^\mp) \cdot\mathcal B^3(K_S^0\to \pi^+\pi^-)}, \phantom{1111111}\\
%	\end{split}
%	\label{fun::bf}
%\end{equation}
\begin{figure}[H]
	\begin{equation}
			{\mathcal B_{\psi(3686)\to \gamma\chi_{cJ}}}\cdot\mathcal{B}_{\chi_{cJ}\to 3K_S^0K^\pm\pi^\mp}= 
				\frac{N^{\rm net}_{\chi_{cJ}}}{N_{\psi(3686)}\cdot\varepsilon \cdot  \mathcal B^3_{K_S^0\to \pi^+\pi^-}},
	\end{equation}	
\label{fun::bf}
\end{figure}

\noindent where $N_{\psi(3686)}$ is the total number of $\psi(3686)$ events in data, $\varepsilon$ is the detection efficiency, and $\mathcal B(K_S^0\to \pi^+\pi^-)$ is
the branching fraction of $K_S^0 \to \pi^+\pi^-$ taken from the PDG~\cite{ref::pdg2024}.
The branching fractions of $\chi_{cJ}\to 3K_S^0 K^{\pm}\pi^{\mp}$ are obtained and summarized in Table \ref{tab:Branchingk-pi+}.

\begin{table*}[htbp]\small
	\caption{The quantities used to calculate the branching fractions of $\chi_{cJ} \to 3K_S^0K^\pm\pi^\mp$, where the first uncertainties are statistical and the second systematic.}
	\centering
%	\resizebox{1\textwidth}{!}{
		\begin{tabular}{l|ccc}
			\hline
			\hline
%			&\multicolumn{3}{c|}{$\chi_{cJ} \to p \bar p K^{0}_{S} K^{\mp} \pi^{\pm}$} \\ \hline
			& $ \chi_{c0}$ & $ \chi_{c1}$ & $ \chi_{c2} $ \\ \hline
			$N^{\rm obs}_{\chi_{cJ}}$ & $343\pm22$ & $1190\pm37$ &$767\pm29$ \\  \hline
			$\varepsilon~(\%)$    & $4.91\pm0.07$ & $5.18\pm0.07$ & $5.22\pm0.07$ \\ \hline
			$\mathcal{B}(\psi(3686)\to\gamma\chi_{cJ})\cdot \mathcal{B}(\chi_{cJ}\to 3K_S^0K^\pm\pi^\mp)~(\times 10^{-6})$ & $7.78\pm0.49\pm0.64$& $25.6\pm0.8\pm1.9$& $16.4\pm0.6\pm1.4$ \\
			\hline
			$\mathcal{B}(\chi_{cJ}\to 3K_S^0K^\pm\pi^\mp)~(\times 10^{-5})$ & $7.95\pm0.50\pm0.65$& $26.2\pm0.8\pm1.9$& $17.2\pm0.7\pm1.5$ \\   \hline
			\hline
		\end{tabular}
%	}
	\label{tab:Branchingk-pi+}
\end{table*}

%\begin{table*}[htbp]
%	\caption{The quantities used for the branching fraction calculations for $\chi_{cJ}\to\bar p \Lambda(1520) K^{0}_{S} \pi^{+}+c.c.$.
%The first uncertainties are statistical and the second systematic.}
%	\centering
%%	\resizebox{1\textwidth}{!}{
%		\begin{tabular}{|l|ccc|}
%			\hline
%			\hline
%%			&\multicolumn{3}{c|}{$\chi_{cJ} \to$ $ \bar p \Lambda(1520) K^{0}_{S} \pi^{+}( p \bar \Lambda(1520) K^{0}_{S} \pi^{-})$}   \\ \hline
%			& $ \chi_{c0}$ & $ \chi_{c1}$ & $ \chi_{c2} $ \\ \hline
%			$N_{\rm sig}$ & $27.0^{+11.4}_{-10.7}$ & $88.2^{+17.3}_{-16.5}$ &$93.8^{+20.1}_{-19.3}$  \\  \hline
%			$\epsilon~(\%)$    & $3.89\pm0.06$ & $5.37\pm0.07$ & $5.60\pm0.07$\\ \hline
%			$\mathcal{B}(\psi(3686)\to\gamma\chi_{cJ})\cdot \mathcal{B}(\chi_{cJ}\to\bar p \Lambda(1520) K^{0}_{S} \pi^{+}+c.c.)~(\times 10^{-6})$ & $1.64^{+0.69}_{-0.65}\pm0.20$ & $3.89^{+0.76}_{-0.73}\pm0.47$ & $3.97^{+0.84}_{-0.82}\pm0.39$ \\   \hline
%			$\mathcal{B}(\chi_{cJ}\to\bar p \Lambda(1520) K^{0}_{S} \pi^{+}+c.c.)~(\times 10^{-5})$ & $1.68^{+0.71}_{-0.66}\pm0.21$ & $3.99^{+0.78}_{-0.75}\pm0.49$ & $4.17^{+0.89}_{-0.86}\pm0.42$ \\   \hline
%			\hline
%		\end{tabular}
%%	}
%	\label{tab:Branchinglambda1520}
%\end{table*}

\section{SYSTEMATIC UNCERTAINTIES}
\label{sec:systematics}

The different sources of systematic uncertainty in the branching fraction measurements are discussed below and summarized in Table~\ref{tab:systematics}.
\begin{table}[htbp]
	\centering
  	\caption{Relative systematic uncertainties (in \%) in the branching fraction measurements.}
 % 	\resizebox{1.0\textwidth}{!}{
  \begin{tabular}{l|ccc}
  	\hline
  	\hline
  	Source & $\phantom{11}\chi_{c0}$\phantom{11} & \phantom{11}$\chi_{c1}$\phantom{11} & \phantom{11}$\chi_{c2}$ \phantom{11} \\
  	\hline
  	$N_{\psi(3686)}$ & $ 0.5$ & $ 0.5$ & $ 0.5$ \\
  	\hline
  	Tracking & $ 2.0$ & $ 2.0$ & $ 2.0$ \\
  	\hline
  	PID      & $ 2.0$ & $ 2.0$ & $ 2.0$ \\
  	\hline
  	$K_{S}^{0}$ reconstruction & $ 4.5$ & $ 4.5$ & $ 4.5$\\
  	\hline
  	$\gamma$ selection & $ 1.0$ & $ 1.0$ & $ 1.0$\\
  	\hline
  	4C kinematic fit & $ 1.0$ & $ 1.9$ & $ 1.3$ \\
  	\hline
  	$M_{3K_S^0K^\pm\pi^\mp}$ fit & $ 5.2$ & $ 3.3$ & $ 4.8$ \\
  	\hline
  	$\mathcal B(K_{S}^{0} \to \pi^{+} \pi^{-})$ & $ 0.2$ & $ 0.2$ & $ 0.2$ \\
  	\hline
  	${\mathcal B}(\psi(3686) \to \gamma \chi_{cJ})$
  	 & $ 2.0$ & $ 2.5$ & $ 2.1$ \\
  	\hline
  	MC statistics & $ 1.6$ & $ 1.5$ & $ 1.4$\\
  	\hline
  	MC generation & $ 0.2$ & $ 0.1$ & $ 0.1$ \\
  	\hline
  	$E^3_\gamma$ factor & $ 1.6$ & $ 1.5$ & $ 3.9$ \\
  	\hline
  	Total & $ 8.2$ & $ 7.4$ & $ 8.7$ \\
  	\hline
  	\hline
  \end{tabular}
%}
  \label{tab:systematics}
\end{table}

The total number of $\psi(3686)$ events in data has been measured to be $N_{\psi(3686)}=(2712.4\pm14.3)\times10^6$ with the inclusive
hadronic data sample, as described in Ref.~\cite{ref::psip-num-inc}. The uncertainty of $N_{\rm \psi(3686)}$ is 0.5\%.

The systematic uncertainties of  tracking and PID efficiencies for $K^{\pm}$ and $\pi^{\pm}$ are estimated with the control samples $J/\psi\to K^*(892)\bar{K}$ and $J/\psi \to p\bar p \pi^+\pi^-$, respectively. They are assigned as 1.0\% for each track, excluding those from $K_S^0$ decays~\cite{ref::tracking}.

The systematic uncertainty of $K^0_S$ reconstruction, including tracking efficiency, $K^0_S$ mass window, and vertex fitters quality, has been estimated by using the control samples $J/\psi \rightarrow K^*(892)^{\pm} ¯\bar{K}^\mp$ and $J/\psi \rightarrow \phi K^*(892)^{\pm} ¯K^\mp$. The systematic uncertainty of $K^0_S$ reconstruction is assigned to be 1.5\% per $K_S^0$~\cite{ref::kso}.

The systematic uncertainty in photon detection is assigned as  1.0\% per photon by studies of the control sample $J/\psi\to\pi^+\pi^-\pi^0$~\cite{ref::gamma-recon}.

To estimate the systematic uncertainties of the MC generation of the $\chi_{cJ}\to 3K_S^0K^\pm\pi^\mp$ decays, we compare our nominal signal efficiencies with those determined from the mixed MC events after varying the relative fractions of the sub-resonant decays by $\pm 1 \sigma$.
The relative changes of the signal efficiencies, 0.2\%, 0.1\% and 0.1\%, are assigned as the  systematic uncertainties for \chic{0}, \chic{1} and \chic{2} decays, respectively.

The systematic uncertainties of the fit to the $M_{3K_S^0K^\pm\pi^\mp}$ distribution are considered below.
The uncertainties due to the signal shape are estimated by changing the signal shape from a Breit-Wigner function convolved with a Gaussian function to the one from signal MC convolved with a Gaussian function. The uncertainties due to the fixed widths of the $\chi_{cJ}$ mesons are estimated by varying the fixed world average values by $\pm1\sigma$. The uncertainties due to the background shape are estimated by varying the nominal ones from the first-order to second-order Chebyshev polynomial functions.
%For the 2D simultaneous fit, we change the background shape in $M_{pK^-}$ or $M_{\bar p K^+}$ from the 1st to a 2nd order Chebyshev polynomial function.
%% and the upper background of $p \bar p K^{0}_{S} K^{\mp} \pi^{\pm}$ .
%The uncertainties from the signal shape are estimated by using the simulated shapes of $M_{p \bar p K^{0}_{S} K^{\mp} \pi^{\pm}}$ and $M_{pK^-}$ or $M_{\bar p K^+}$ convolved with a Gaussian function.
%while the parameters of the smeared Gaussian function are fixed at those obtained from the 1D fits for the 2D simultaneous fit.
The uncertainties of the simultaneous fit method are estimated by changing the nominal fit to a one-dimensional fit of the $M_{3K_S^0K^\pm\pi^\mp}$ distribution, with a background shape derived from data events in the three-dimensional $K_S^0$ sidebands. The uncertainties of the fit range are estimated by changing them from $[3.31,3.61]$ GeV/$c^2$ to $[3.28,3.58]$ GeV/$c^2$ and $[3.35,3.65]$ GeV/$c^2$.
Adding these uncertainties in quadrature gives the total  systematic uncertainties due to the $M_{3K_S^0K^\pm\pi\mp}$ fit to be 5.2\%, 3.3\%, and 4.8\%
for $\chi_{c0}, \chi_{c1}$ and $\chi_{c2}$, respectively.

%The systematic uncertainties due to the fit bias are estimated by performing input-output checks based on inclusive MC sample.
%The differences between the measured and input branching fractions are taken as the systematic uncertainties,
%which are 3\%,~3\% and 3\% for  $\chi_{c0,1,2}\to p \bar p K^{0}_{S} K^{\mp} \pi^{\pm}$; and
%10\%, 10\% and 7\% for $\chi_{c0,1,2}\to \bar p \Lambda(1520) K^{0}_{S}  \pi^{+}+c.c.$,~respectively.
%
%The systematic uncertainties from the low end of the fit range are estimated by using the alternative fit ranges of
%$[3.25,3.6.0]$ and $[3.35,3.6]$ GeV$/c^{2}$.
%The larger differences of the fitted signal yields are taken as the systematic uncertainties, which are
%4.9\%, 1.5\% and 0.8\% for $\chi_{c0,1,2}\to p \bar p K^{0}_{S} K^{\mp} \pi^{\pm}$;
%and 2.6\%, 0.2\%, and 0.1\% for $\chi_{c0,1,2}\to \bar p \Lambda(1520) K^{0}_{S}  \pi^{+}+c.c.$,~respectively.

The systematic uncertainties from the 4C kinematic fit is estimated by comparing the signal efficiencies after and before the helix parameter correction, with parameters taken from Ref.~\cite{ref::helixp}.
The changes of the signal efficiencies are assigned as the systematic uncertainties, which are 1.0\%, 1.9\% and 1.3\% for $\chi_{c0}, \chi_{c1}$ and $\chi_{c2}$.

The systematic uncertainties due to MC statistics
are 1.6\%, 1.5\%, and 1.4\% for $\chi_{c0}, \chi_{c1}$ and $\chi_{c2}$, respectively.

The uncertainties of the branching fractions quoted from the PDG~\cite{ref::pdg2024} are 2.0\%, 2.5\%, and 2.1\% for $\psi(3686)\to \gamma\chi_{c0,1,2}$ and
0.07\% for $K^{0}_{S} \to \pi^+ \pi^-$.

The systematic uncertainties in the $E_\gamma^3 $ factor, which is used to take into account the electromagnetic transition rates of the lowest multipoles, are
%considered below.
%The uncertainties due to signal fit are
estimated by varying the signal shape from a Breit-Wigner function convolved with a Gaussian function to a Breit-Wigner function multiplied by the $E_\gamma^3$ factor~\cite{Anashin:2010dh}. The differences in the signal yields are assigned as the systematic uncertainties, which are 1.5\%, 1.3\%, and 3.9\% for $\chi_{c0}, \chi_{c1}$ and $\chi_{c2}$.
%The uncertainties due to detection efficiencies are estimated by comparing with the detection efficiencies of signal MC events with and without this factor.

%The differences in the signal efficiencies are taken from these two sources in quadrature,

For each signal decay, the total systematic uncertainty is obtained by adding all systematic uncertainties in quadrature.

\section{Summary}
By analyzing $(2712.4\pm14.3)\times10^6$ $\psi(3686)$ events collected with the BESIII detector operating at the BEPCII collider,
 the decays  $\chi_{c0,1,2} \to 3K_S^0K^\pm\pi^\mp$ are observed for the first time with statistical significances greater than $10\sigma$.
Their decay branching fractions are determined to be
$$
\begin{aligned}
	\mathcal{B}(\chi_{c0}\to 3K_S^0K^\pm\pi^\mp) = (7.95\pm0.50\pm0.65)\times10^{-5}, \\
	\nonumber
	\mathcal{B}(\chi_{c1}\to 3K_S^0K^\pm\pi^\mp) = (2.52\pm0.08\pm0.19)\times10^{-4}, \\
	\nonumber
	\mathcal{B}(\chi_{c2}\to 3K_S^0K^\pm\pi^\mp) = (1.72\pm0.07\pm0.15)\times10^{-4}. \\
	\nonumber
\end{aligned}
$$
After considering the intermediate states and the effects of phase space, these results are comparable with the branching fractions in Table~\ref{tab::reledecay}. Through the measurement of the branching fractions of $\chi_{cJ}\to 3K_S^0K^\pm\pi^\mp$, it is possible to derive valuable information to deepen our understanding of the mechanisms of $\chi_{cJ}$ decays.

\section{ACKNOWLEDGMENTS}
% 2024-10-17
The BESIII Collaboration thanks the staff of BEPCII and the IHEP computing center for their strong support. This work is supported in part by National Key R\&D Program of China under Contracts Nos. 2020YFA0406300, 2020YFA0406400, 2023YFA1606000; National Natural Science Foundation of China (NSFC) under Contracts Nos. 12035009, 11635010, 11735014, 11935015, 11935016, 11935018, 12025502, 12035013, 12061131003, 12192260, 12192261, 12192262, 12192263, 12192264, 12192265, 12221005, 12225509, 12235017, 12361141819; the Chinese Academy of Sciences (CAS) Large-Scale Scientific Facility Program; the CAS Center for Excellence in Particle Physics (CCEPP); Joint Large-Scale Scientific Facility Funds of the NSFC and CAS under Contract No. U1832207; CAS under Contract No. YSBR-101; 100 Talents Program of CAS; The Institute of Nuclear and Particle Physics (INPAC) and Shanghai Key Laboratory for Particle Physics and Cosmology; Agencia Nacional de Investigaciny Desarrollo de Chile (ANID), Chile under Contract No. ANID PIA/APOYO AFB230003; German Research Foundation DFG under Contract No. FOR5327; Istituto Nazionale di Fisica Nucleare, Italy; Knut and Alice Wallenberg Foundation under Contracts Nos. 2021.0174, 2021.0299; Ministry of Development of Turkey under Contract No. DPT2006K-120470; National Research Foundation of Korea under Contract No. NRF-2022R1A2C1092335; National Science and Technology fund of Mongolia; National Science Research and Innovation Fund (NSRF) via the Program Management Unit for Human Resources \& Institutional Development, Research and Innovation of Thailand under Contracts Nos. B16F640076, B50G670107; Polish National Science Centre under Contract No. 2019/35/O/ST2/02907; Swedish Research Council under Contract No. 2019.04595; The Swedish Foundation for International Cooperation in Research and Higher Education under Contract No. CH2018-7756; U. S. Department of Energy under Contract No. DE-FG02-05ER41374.


\begin{thebibliography}{99}

\bibitem{bes3-white-paper} M. Ablikim {\it et al.} (BESIII Collaboration),
\href{https://iopscience.iop.org/article/10.1088/1674-1137/44/4/040001}
{Chin. Phys. C {\bf 44}, 040001 (2020).}

\bibitem{qcd1} W. Kwong, J. L. Rosner and C. Quigg,
\href{https://inspirehep.net/files/92cd561c57c96570c626e50005a0e44e}
{Ann. Rev. Nucl. Part. Sci. {\bf37}, 325 (1987).}

\bibitem{qcd2} E. Eichten, S. Godfrey, H. Mahlke and J. L. Rosner,
\href{https://journals.aps.org/rmp/pdf/10.1103/RevModPhys.80.1161}
{Rev. Mod. Phys. {\bf80}, 1161 (2008).}

\bibitem{ref::pdg2024} S. Navas \textit{et al.} (Particle Data Group),
\href{https://pdg.lbl.gov/2024/download/PhysRevD.110.030001.pdf}{Phys. Rev D \textbf{110}, 030001 (2024).}

\bibitem{2004kpm}
N.~Brambilla \textit{et al.} (Quarkonium Working Group), \href{https://doi:10.5170/CERN-2005-005}{CERN Yellow Reports: Monographs, CERN-2005-005}.
\bibitem{2010cs} H. W. Huang and K. T. Chao, \href{https://doi:10.1140/epjc/s10052-010-1534-9}{Eur. Phys. J. C \textbf{71}, 1534 (2011)}.
%\bibitem{COM3}A. Petrelli, \href{https://doi.org/10.1016/0370-2693(96)00459-5}{Phys. Lett. B \textbf{380}, 159 (1996)}.
%\bibitem{COM4}J. Bolz, P. Kroll, and G. A. Schuler, \href{https://doi.org/10.1007/s100529800716}{Eur. Phys. J. C \textbf{2}, 705 (1998)}.
%\bibitem{COM5}S. H. M. Wong, \href{https://doi.org/10.1007/s100520000376}{Eur. Phys. J. C \textbf{14}, 643 (2000)}.
\bibitem{ref::kkpiphi} M.~Ablikim \textit{et al.} (BESIII Collaboration), \href{https://doi:10.1103/PhysRevD.91.112008}{Phys. Rev. D \textbf{91}, 112008 (2015).}
\bibitem{ref::psip-num-inc} M. Ablikim {\it et al.} (BESIII Collaboration),
\href{https://arxiv.org/pdf/2403.06766}{Chin.Phys.C \textbf{48}, 093001(2024)}.

\bibitem{ref::detector} M. Ablikim {\it et al.} (BESIII Collaboration), \href{https://www.sciencedirect.com/science/article/pii/S0168900209023870}{Nucl. Instrum. Methods Phys. Res., A {\bf614}, 345 (2010).}	
\bibitem{ref::collider} C.~H.~Yu {\it et al.}, \href{https://accelconf.web.cern.ch/ipac2016/doi/JACoW-IPAC2016-TUYA01.html}{Proceeding of IPAC2016, Busan, Korea, 2016}
\bibitem{Tof1} X.~Li {\it et al.}, \href{https://link.springer.com/article/10.1007/s41605-017-0014-2} {Radiat. Detect. Technol. Methods {\bf 1}, 13 (2017).}

\bibitem{Tof2} Y.~X.~Guo {\it et al.}, \href{https://link.springer.com/article/10.1007/s41605-017-0012-4} {Radiat. Detect. Technol. Methods {\bf 1}, 15 (2017).}

\bibitem{Tof3} P.~Cao {\it et al.}, \href{https://www.sciencedirect.com/science/article/pii/S0168900219314068} {Nucl. Instrum. Meth. A {\bf 953}, 163053 (2020).}

\bibitem{Geant4} S.~Agostinelli \textit{et al.} ({\sc geant4} Collaboration), \href{https://www.sciencedirect.com/science/article/pii/S0168900203013688?via\%3Dihub}{Nucl. Instrum. Methods Phys. Res. {A \bf506}, 250 (2003).}
\bibitem{Jadach01} S.~Jadach, B.~F.~L.~Ward~and Z.~Was, \href{https://journals.aps.org/prd/abstract/10.1103/PhysRevD.63.113009}{Phys. Rev. {D \bf63}, 113009 (2001).}
\bibitem{Lange01} D.~J.~Lange,  \href{https://www.sciencedirect.com/science/article/pii/S0168900201000894?via\%3Dihub}{Nucl. Instrum. Methods Phys. Res. {A \bf462}, 152 (2001).}
\bibitem{Lange02}
 R.~G.~Ping, \href{https://iopscience.iop.org/article/10.1088/1674-1137/32/8/001}{Chin. Phys. {C \bf32}, 599 (2008).}

\bibitem{Lundcharm00} J.~C.~Chen, G.~S.~Huang, X.~R.~Qi, D.~H.~Zhang and Y.~S.~Zhu, \href{https://journals.aps.org/prd/abstract/10.1103/PhysRevD.62.034003}{Phys. Rev. {D \bf62}, 034003 (2000).}

\bibitem{PHOTOS} E.~Richter~Was, \href{https://www.sciencedirect.com/science/article/pii/037026939390062M}{Phys. Lett. {B \bf303}, 163 (1993).}
\bibitem{topoana} M.~Ablikim {\it et al.} (BESIII Collaboration), \href{https://doi.org/10.1016/j.cpc.2020.107540}{Computer Physics Communications{ \bf258}, 107540 (2021).}
\bibitem{lum} M.~Ablikim {\it et al.} (BESIII Collaboration), \href{https://iopscience.iop.org/article/10.1088/1674-1137/37/12/123001}{Chin. Phys. {C \bf37}, 123001 (2013).}
\bibitem{ref::generate} W.~M.~Tanenbaum {\it et al.}, \href{https://journals.aps.org/prd/abstract/10.1103/PhysRevD.17.1731}{Phys. Rev. {D \bf17}, 1731 (1978).}
\bibitem{ref::tracking} M.~Ablikim {\it et al.} (BESIII Collaboration), \href{https://journals.aps.org/prd/pdf/10.1103/PhysRevD.83.112005}{Phys. Rev D \textbf{83}, 112005(2011)}.
\bibitem{ref::kso} M.~Ablikim {\it et al.} (BESIII Collaboration).
\href{https://journals.aps.org/prd/abstract/10.1103/PhysRevD.92.112008}{Phys. Rev. D \textbf{92},
	112008 (2015)}.
\bibitem{ref::gamma-recon} M. Ablikim \textit{et al}. (BESIII Collaboration).
\href{https://journals.aps.org/prd/pdf/10.1103/PhysRevD.83.112005}{Phys. Rev D \textbf{83}, 112005(2011)}.
\bibitem{ref::helixp} M.~Ablikim {\it et al.} (BESIII Collaboration), \href{https://doi.org/10.1103/PhysRevD.87.012002}{Phys. Rev. {D \bf87}, 012002 (2013).}
%\bibitem{ref::kkk0k0} M. Ablikim {\it et al.} (BESIII Collaboration), \href{https://www.sciencedirect.com/science/article/pii/S037026930501378X?via\%3Dihub}{Phys. Lett. {B \bf630}, 21 (2005).}
\bibitem{Anashin:2010dh}
V.~V.~Anashin, \textit{et al.}
%``Measurement of $\mathcal{B}(J/\psi \to \eta_c \gamma)$ at KEDR,''
\href{https://doi.org/10.1142/S2010194511000791}{Int. J. Mod. Phys. Conf. Ser. \textbf{02}, 188-192 (2011)}.


\end{thebibliography}
\end{document}